\documentclass[%
 reprint,
 amsmath,amssymb,
 aps,
floatfix,
]{revtex4-2}

\usepackage{graphicx}
\usepackage{dcolumn}
\usepackage{bm}
\usepackage{hyperref}

\begin{document}

\preprint{APS/123-QED}

\title{\Large{Local Changes in Protein Filament Properties
Drive Large-Scale Membrane Transformations Involved in
Endosome Tethering and Fusion}}

\author{Ashesh Ghosh$^{1,}$}
\altaffiliation[Now at ]{Department of Chemistry and Chemical Engineering, University of California Berkeley, CA 94720}
\author{Andrew J. Spkaowitz$^{1,2,3,}$}%
\email{ajspakow@stanford.edu}
\affiliation{%
 $^1$Department of Chemical Engineering, Stanford University, Stanford, CA 94305 \\
$^2$Biophysics Program, Stanford University, Stanford, CA 94305 \\
$^3$Department of Materials Science \& Engineering, Stanford University, Stanford, CA 94305
}%

\date{\today}

\begin{abstract}
Large-scale cellular transformations are triggered by subtle physical and structural changes in individual
biomacromolecular and membrane components.
A prototypical example of such an event is the orchestrated fusion of membranes within an endosome that enables
transport of cargo and processing of biochemical moieties.
In this work, we demonstrate how protein filaments on the endosomal membrane surface can leverage a rigid-to-flexible
transformation to elicit a large-scale change in membrane flexibility to enable membrane fusion.
We develop a polymer field-theoretic model that captures molecular alignment arising from nematic interactions
with varying surface density and fraction of flexible
filaments, which are biologically controlled within the endosomal membrane.
We then predict the collective elasticity of the filament brush in response to changes in the filament alignment,
predicting a greater than 20-fold increase of the effective membrane elasticity over the bare membrane elasticity
that is triggered by filament alignment.
These results show that the endosome can modulate the filament properties to orchestrate 
membrane fluidization that facilitates vesicle fusion, providing an example of 
how active processes that modulate local molecular properties can 
result in large-scale transformations that are essential to cellular survival.
\end{abstract}

\maketitle


\begin{figure}[t]
    \centering
    \includegraphics[width=0.45\textwidth]{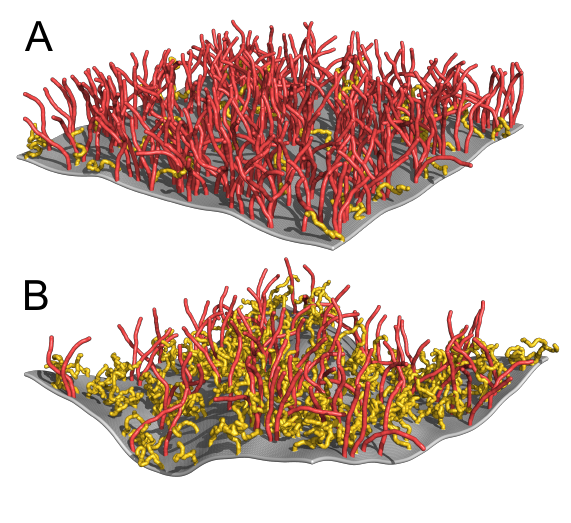}
    \caption{Schematic of an endosomal membrane with EEA1 brush layer. The A and B images show the collective behavior of the brush layer in the predominantly extended and predominantly collapsed states on the surface of a patch of elastic membrane. Yellow and Red colored polymers are in the collapsed flexible and extended rigid conformations, respectively. The effective elastic rigidity of the membrane and brush is higher in the aligned state of polymer brush layer (panel A). Molecular-level changes that transform polymers from extended to flexible conformation (red $\to$ yellow) alter membrane rigidity and induce large length-scale fluctuations.}
    \label{fig:schematic}
\end{figure}

Cellular self-organization and large-scale orchestrated transformations
are driven by coordinated 
biophysical, mechanical, and dynamical 
processes involving numerous biomacromolecules and biological membranes.
A fundamental challenge in understanding and predicting cellular self-organization and dynamics stems from
the need to determine how molecular components (\emph{e.g.} RNA, proteins, lipids) leverage subtle physical changes to drive 
dramatic cellular transformations across different length and time scales~\cite{mccusker2020cellular,dye2021self,recouvreux2016molecular,furse2018lipids}.
The complex interplay of microscopic chemical events in membranes, including lipid raft formation~\cite{ikonen2001roles}, coupled lipid-protein interactions~\cite{brown2017soft}, dynamin-driven membrane fission~\cite{pannuzzo2018role}, and active transport of fluid across membranes~\cite{girard2005passive,wilhelm2008out}, determine critical biological fates such as cell shape and function~\cite{bassereau20182018}.
Coordinated membrane fusion enables the transport of critical 
cargo and biochemicals to and from the cell~\cite{lipatova2015ypt}, which is an essential biological process involving subtle biophysical
events.

In this work, we develop a physical model for the physical transformations in the endosomal membrane that
underlie membrane fusion and transport, which 
serves as a prototypical example of cellular self-organization and dynamics.
Recently, allosteric changes to the early endosome tether protein EEA1 (Fig.~\ref{fig:schematic}) has been shown to play an important role in mechanical pulling of membranes to close spatial proximity upon binding with vesicle bearing small GTPase Rab5 in its GTP bound form Rab5(GTP).
EEA1 is a coiled-coil dimeric molecule with a contour length of $L \!\!\sim\!\! 222 \pm 26$~nm and a persistence length of $l_p^{\text{EEA1}}\!\!\sim\!\! 246\pm42$~nm.
EEA1 predominantly exists in ``extended'' or rigid conformation (Fig.~\ref{fig:schematic}A) in its free N-terminus unbound form~\cite{murray2016endosomal}. 
Upon binding to Rab5(GTP) the contour length essentially remains the same.
However, the molecule shifts to a more ``flexible'' conformation (Fig.~\ref{fig:schematic}B) with an average end-to-end distance of $\sim\!\! 122\pm 50$~nm,
coinciding 
with a persistence length of $l_p^{\text{EEA1(GTP)}}\!\!\sim\!\! 74\pm 3$~nm~\cite{murray2016endosomal}.

Changes in the population density of the structurally rigid (or flexible) EEA1 filaments is thought to play a crucial role in mechanical pulling of membranes via thermal and active enzymatic fluctuations present in the cellular media.
Moreover, EEA1 is present in high density patches on the membrane surface and thus forms a switchable polymer brush layer~\cite{singh2023two}.
It is to be noted that such a structure-driven entropic transition is not unique to EEA1-tether system, but common in coil-coiled tether proteins, such as GCC185 that binds to Rab9(GTP)~\cite{cheung2015protein}.
Hence, these long coil-coiled tethers can act as a mechanical switch in bringing distant membranes to physical proximity to facilitate fusion~\cite{murray2016endosomal,wilson2000eea1}.

We provide a theoretical model and explanation of how the collective brush layer alters the effective elasticity of the membrane and makes it more susceptible for fusion.
More concretely, we construct a general thermodynamic model of two types of semiflexible polymers in a solvent and specialize to our case to determine how nematic interactions that originate from hydrophobic moieties along the chain backbone of coil-coiled proteins 
dictate chain alignment and alter the collective membrane rigidity.
This work provides fundamental insight into how the cell can control molecular properties of specific biomacromolecules to elicit
large-scale transformations involved in major biological events.


We consider an incompressible polymer solution of total volume $V$, with $n_{s}$ solvent molecules with molecular volume $v_{s}$, $n_{p}^R$ rigid polymer chains with contour length $L_R$ and cross-sectional area $A_R$ and $n_{p}^F$ flexible polymer chains with contour length $L_F$ and cross-sectional area $A_F$.
The polymer chains are modeled using the wormlike chain model, which describes the polymer chains as inextensible elastic threads that are subjected to thermal fluctuations~\cite{kratky1949rontgenuntersuchung,saito1967statistical}.
The polymer configuration of the $i$th polymer of type $\alpha \in [R,F]$ at arc-length position $s$ (note, $s\!=\!0$ at one end and $s\!=\!L_{\alpha}$ at the opposite end) 
is defined by the space curve $\vec{r}_{i}^\alpha(s)$, and the local tangent vector $\vec{u}_{i}^\alpha(s) = \partial \vec{r}_{i}^\alpha(s)/\partial s$
gives the tangent orientation of the monomer segment at $s$.
Inextensibility is strictly enforced by ensuring that $|\vec{u}_{i}^\alpha(s)| = 1$ for all configurations of the system.
The position of the $j$th solvent molecule is given by $\vec{r}_{j}$.
The bending rigidity of the semiflexible polymers are given by their respective persistence lengths $l_{p}^\alpha$,
and we also express the chain length in dimensionless units as
the number of Kuhn lengths $N_\alpha = L_\alpha/(2 l_{p}^\alpha)$,
where the Kuhn length $b_\alpha = 2 l_{p}^\alpha$ gives the statistical segment length of a polymer.

The system energy includes contributions for polymer bending deformation, solvent-polymer mixing enthalpy 
captured through effective Flory-Huggins $\chi$-parameters~\cite{fredrickson2006equilibrium}, and nematic alignment free energy of polymer chains through an effective Maier-Saupe interaction~\cite{maier1958simple,maier1959einfache}. 
The total energy is given by
\begin{flalign}
        \beta E &= \displaystyle \sum_{\alpha\in \{R,F\}} \Bigg\{ 
        \sum_{i=1}^{n_p^\alpha} \frac{l_p^\alpha}{2} \int_0^{L_{\alpha}} ds \left(\frac{\partial \vec{u}_i^{\alpha}(s)}{\partial s}\right)^2 \nonumber \\ 
        \! & +\chi_\alpha \int d\Vec{r}\, \hat{\phi}_s(\vec{r})\hat{\phi}_p^\alpha(\vec{r})
        -\frac{a_\alpha}{2}\int d\Vec{r}\, \hat{\mathbf{S}}_\alpha(\vec{r}):\hat{\mathbf{S}}_\alpha(\vec{r}) \Bigg\} \nonumber \\
        & +\chi_{RF} \int d\Vec{r}\, \hat{\phi}_p^R(\vec{r})\hat{\phi}_p^F(\vec{r})
        -a_{RF} \int d\Vec{r}\, \hat{\mathbf{S}}_R(\vec{r}):\hat{\mathbf{S}}_F(\vec{r}),
        \label{Eq.energy}
\end{flalign}
\noindent where $\hat{\phi}_{s}(\vec{r})$ and $\hat{\phi}_{p}^\alpha(\vec{r})$ respectively define the local dimensionless
density (volume fraction) of the solvent and polymer molecules of type $\alpha$ at spatial location $\vec{r}$, and $\hat{\mathbf{S}}_\alpha(\vec{r})$
is the tensorial nematic order parameter density of state $\alpha$ at $\vec{r}$~\cite{ghosh2022semiflexible,doi1988theory,de1993physics}.
Defining a generalized volume fraction in terms of real spherical harmonics~\cite{blanco1997evaluation} allows us to write the thermodynamic energy in a consolidated form as done in Ref.~\cite{ghosh2022semiflexible} for a one-component polymer system (details in Supplemental Materials, Secs.~I-IV).
Writing the canonical partition function followed by a series of standard polymer field-theoretic transformations~\cite{fredrickson2006equilibrium} and 
using the saddle-point approximation~\cite{fredrickson2006equilibrium}, we  write the homogeneous mean-field free energy density of the system as
\begin{flalign}
    \beta f_0 & = \frac{1-\phi_p}{v_s}\ln{(1-\phi_p)}+\sum_{\alpha\in \{R,F\}} \Bigg\{\frac{\phi_p^\alpha}{L_\alpha A_\alpha}\ln{(\phi_p^\alpha)}\nonumber \\
    & + \chi_\alpha \phi_p^\alpha(1-\phi_p)- \frac{a_\alpha}{3} ({\phi}_p^{\alpha})^2 m_\alpha^2 +  \frac{\phi_p^\alpha}{L_\alpha A_\alpha}\ln{(q_\alpha)}\Bigg\} \nonumber \\
    & + \chi_{RF} \phi_p^R \phi_p^F - \frac{2a_{RF}}{3}{\phi}_p^{R}{\phi}_p^{F}m_Rm_F,
    \label{Eq.mean_field}
\end{flalign}
\noindent where $\phi_p^\alpha$ is the bulk volume fraction of polymer $\alpha$, 
and the total polymer volume fraction is $\displaystyle \phi_p=\phi_p^R + \phi_p^F$. The nematic order parameter $m_\alpha$ of polymer $ \alpha$ is defined as 
$\displaystyle m_\alpha  =\frac{1}{L_\alpha} \int_0^{L_\alpha} ds \left\langle P_2(u_z^\alpha(s)) \right\rangle_{0_\alpha}$~\footnote{$\displaystyle P_2(x)=(3x^2-1)/2$ is the Legendre polynomial of order 2.}.
The average $\langle \dots \rangle_{0_\alpha}$ is taken with respect to the single-chain mean field energy, defined by
\begin{flalign}
     \beta \mathcal{E}_0^\alpha =\int_0^{L_\alpha} ds \left\{ \frac{l_p^\alpha}{2}\left(\frac{\partial \vec{u}^\alpha(s) }{\partial s}\right)^{\! 2} - 
     \gamma_\alpha \left[ (u_z^\alpha(s))^2-\frac{1}{3}\right] \right\},
     \label{Eq.single_energy}
\end{flalign}
where $\gamma_\alpha =  (a_\alpha \phi_p^\alpha m_\alpha A_{\alpha}+a_{RF} \phi_p^\beta m_\beta A_\beta)$ gives the molecular field strength
for the $\alpha$-type polymer. 
The single-chain partition function $q_\alpha$ of polymer $\alpha$ is evaluated from a Boltzmann-weighted
sum over all polymer-chain conformations with respect to the mean-field energy $\beta \mathcal{E}_0^\alpha$.
We note that in the absence of any molecular quadrupole field interactions ($a\to 0$ and $q\to 1$ limit), Eq.~\ref{Eq.mean_field} reduces to a
multicomponent Flory-Huggins (FH) mean-field theory~\cite{fredrickson2006equilibrium}.
This FH free energy corresponds to the isotropic state of the multicomponent system.

In this work, we assume the volume fractions in the isotropic and nematic states are the same in order to model transitions
in the filament brush for a fixed membrane surface coverage, and we assume $a_{R} = a_{F} = a_{RF}=a$.
With these assumptions, we write the Helmholtz free energy of the nematic phase relative to the isotropic phase for the alignment strength
$\hat{a}=a\phi_p L A$ to be
\begin{flalign}
    \beta\Delta f_0 
    = \frac{\hat{a}}{3}\left(f_Rm_R+f_Fm_F\right)^2 - f_R\ln{q_R} - f_F\ln{q_F},
    \label{Eq.free_energy}
\end{flalign}
\noindent where $f_\alpha = \phi_p^\alpha/\phi_p$.
The Maier-Saupe parameter $\hat{a}$ should be understood as altering the 
surface coverage of filaments on the membrane.
The stability of the nematic phase is determined by finding the curvature of $\beta \Delta f_0$ with respect to the overall variational parameter, defined through the relation $\langle m \rangle = f_Rm_R+f_Fm_F$.
Finally, we note that the end-to-end distance can be calculated from the tangent correlation functions, similar to Ref.~\cite{spakowitz2003semiflexible,ghosh2022semiflexible}. 

Fluctuations around the mean-field state~\cite{fredrickson2006equilibrium} of the polymer brush layer predict 
the Frank elastic energies of the system that contribute to the effective membrane bending rigidity.
Frank elastic energies provide energetic contributions of the polymer layer with respect to the aligned state 
in terms of the normal modes of deformation such as bend, twist, and splay~\cite{frank1958liquid,priest1973theory,ghosh2022semiflexible}. 
Considering the Canham-Helfrich Hamiltonian~\cite{helfrich1973elastic,canham1970minimum} using a Monge-Gauge
representation~\cite{mazharimousavi2017generalized},
we note the following relation in the presence of a polymer brush layer (for details see, Supplemental Materials, Sec.~V):
\begin{eqnarray}
    \kappa_{\text{mem}}^{\text{eff}} = \kappa_{\text{mem}} + K_{\text{splay}}\sqrt{R_\parallel^2} \ge \kappa_{\text{mem}},
    \label{Eq.membrane_kappa}
\end{eqnarray}
where only the splay modulus $K_{\text{splay}}$ is seen to contribute to altering the effective
membrane bending modulus, and $R_\parallel^2$ denotes the mean-squared end-to-end distance of the polymer along the 
nematic alignment direction.
Previous work~\cite{terzi2019consistent} determines the membrane elasticity to be coupled to the lipid-bilayer splay modulus, which in spirit is similar to what we obtain. 
However, our single layer membrane elasticity is coupled to splaying of the outer protein-brush layer, and the nematic state of the brush layer becomes crucial for determining the effective membrane rigidity.
The stretching modulus of the membrane remains unchanged, since 
modes of deformation that 
are affected by the brush layer are associated with directions that are predominantly perpendicular to the in-plane stretching modes. 
For our case of two different polymers, the effective bending modulus is determined 
to be the population average $\sim K_{\text{splay}}\sqrt{R_\parallel^2}$.
Details of a microscopic theory of how fluctuations around the mean-field solution can be used to calculate Frank-elastic constants for semiflexible polymers 
is given in Ref.~\cite{ghosh2022semiflexible}.
Equation~\ref{Eq.membrane_kappa} suggests that when $K_{\text{splay}}\!=\!0$ in the absence of nematic alignment, 
the effective membrane elasticity comes from internal rigidity of the lipid bilayer.
Hence, thermal (and perhaps active) fluctuations present in the system result in 
larger height fluctuations of the membrane~\cite{takatori2020active} (\emph{cf.} Fig.~\ref{fig:schematic}B). 
Under conditions of alignment of the polymer molecules, $K_{\text{splay}}\! > \!0$, and effective membrane elasticity 
increases.
Physically, this corresponds to suppressed height fluctuations~\cite{sapp2016suppressing} of the membrane due to the extended configurations within the
polymer brush layer (\emph{cf.} Fig.~\ref{fig:schematic}A).

We show results for the two types of polymers with variable dimensionless chain length $N_\alpha=L/2l_p^\alpha$, 
where the contour length and persistence lengths are set by experimental measurements of EEA1 filaments.
We also set all Maier-Saupe parameters $a_\alpha$ 
to be the same and cross-sectional area is considered to be the same for EEA1 protein in its extended and flexible states.
We note that for a fixed composition (i.e. fractions of rigid and flexible polymers) transitioning to a liquid-crystalline phase, the nematic order parameter for the two polymers exhibit a first-order phase transition.
\begin{figure}[t]
    \centering
    \includegraphics[scale=0.4]{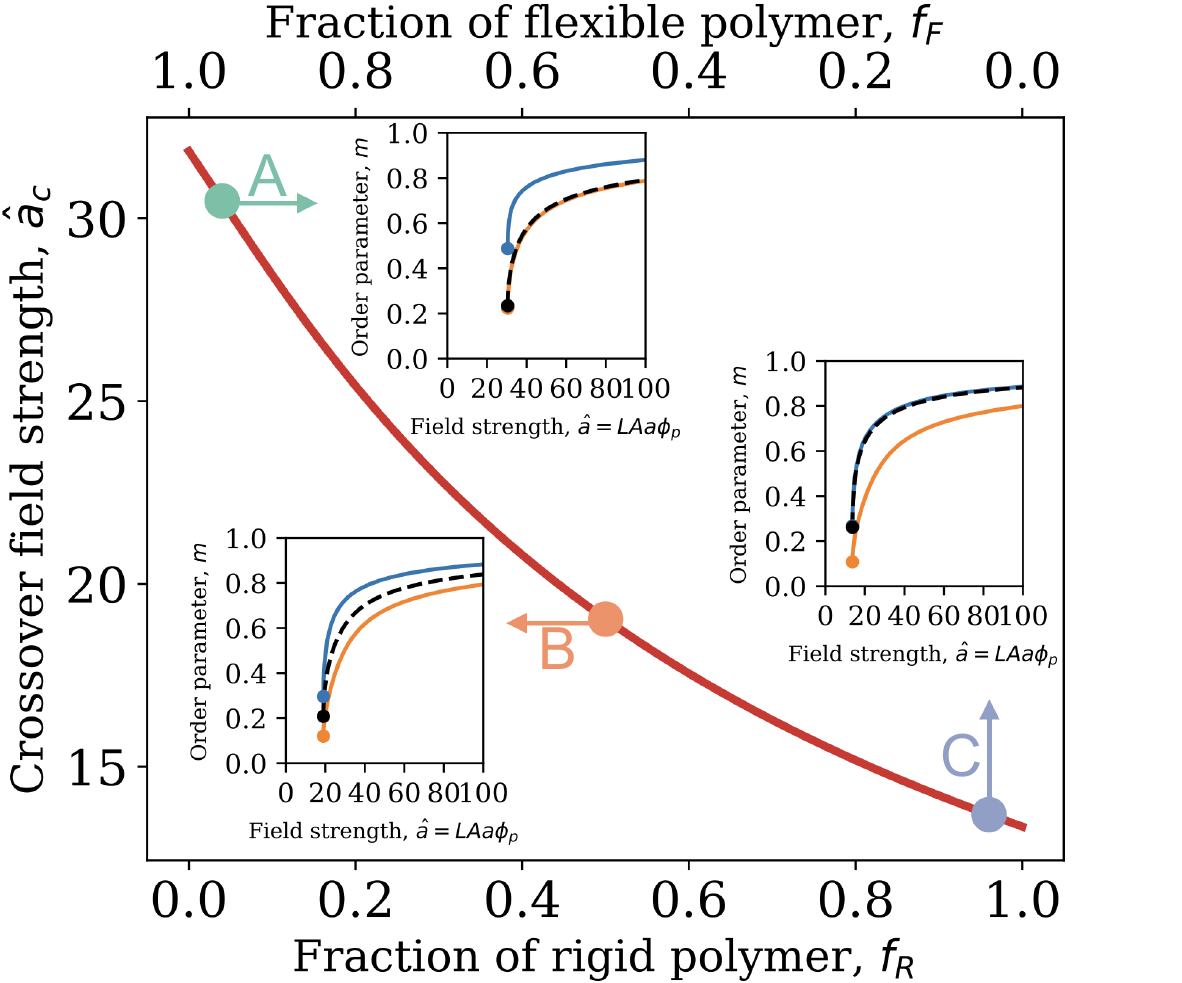}
    \caption{Critical strength of aligning field as a function of the fraction of rigid polymer (lower X-axis) is plotted. The order parameter variation is shown for three states corresponding to $f_R=0.04, 0.50 \text{ and }0.96$ respectively [blue line is $m_R$, orange is $m_F$ and the black dashed line follows $\langle m\rangle$].}
    \label{fig:phase_diag}
\end{figure}

The main plot of Fig.~\ref{fig:phase_diag} shows the field strength 
at the nematic transition $\hat{a}_c$
as a function of fraction of rigid polymers $f_{R}$.
The transition field strength $\hat{a}_c$ 
systematically decreases with increasing $f_R$, indicating the system transitions to a
nematic state at lower effective strength of alignment interactions.
Phase diagrams for the three states [$f_R\!=0.04$~(A), $0.50$~(B), and $0.96$~(C)] showing the
nematic order parameters for the flexible and rigid polymers ($m_{F}$ and $m_{R}$, respectively)
are given in the inset plots of
Fig.~\ref{fig:phase_diag}.
The two colors in the inset figures represent $m_R$ (rigid, blue) and $m_{F}$ (flexible, orange).
The population-weighted average nematic order parameter is shown in black dashed line.
The limit of stability of the nematic phase $\hat{a}_{s}$ is shown as the dots in the inset phase diagrams.
Both the rigid and flexible polymers exist in the isotropic state with $m_\alpha=0$ for $a\!<\!a_s$.
The polymer brush layer exhibits nematic alignment for $\!a\!>\!a_s$, resulting in non-zero nematic ordering for both
rigid and flexible polymers.

\begin{figure}[t]
    \centering
    \includegraphics[scale=0.4]{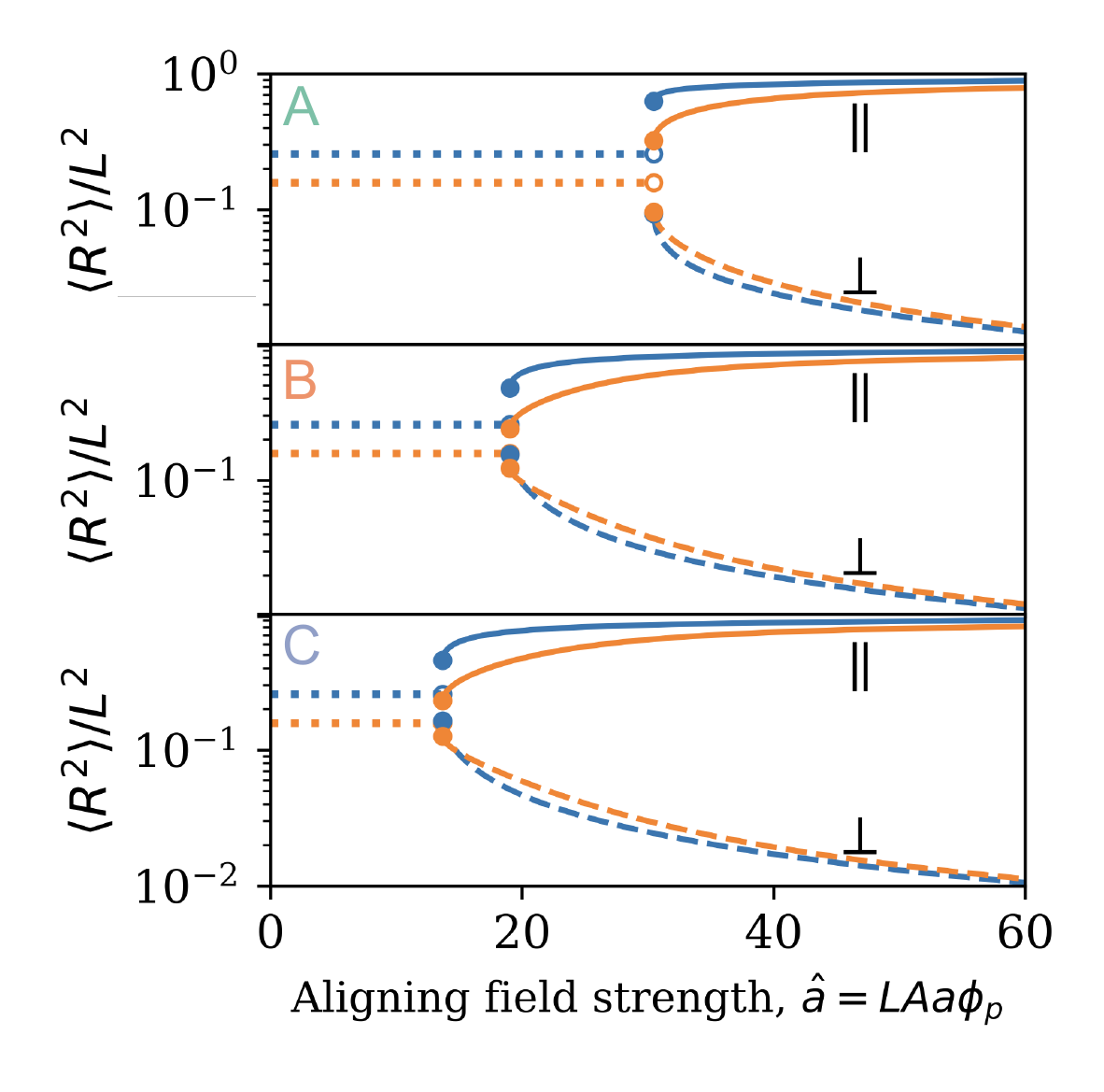}
    \caption{Mean-squared end-to-end distance as a function of aligning field strength for the three states in Fig.~\ref{fig:phase_diag}. Dotted line represents an isotropic state and solid and dashed lines represent $\langle R_\parallel^2\rangle$ (along the aligning field) and $\langle R_\perp^2\rangle$ ($\perp^{\text{r}}$ to the aligning field) respectively [blue $\equiv \text{rigid}$ and orange $\equiv \text{flexible}$].}
    \label{fig:end_2_end}
\end{figure}

The isotropic state of the system exhibits orientational symmetry, such that the mean-squared end-to-end distance has equivalent contributions from all directions~\cite{spakowitz2003semiflexible,ghosh2022semiflexible}.
Upon transitioning to the nematic state, the broken symmetry of correlation lengths along parallel and perpendicular to the field directions result in an `elongated' polymer along the nematic-director direction~\cite{ghosh2022semiflexible} (i.e. perpendicular to the membrane).
We show the end-to-end distance as a function of aligning field strength for three representative states in Fig.~\ref{fig:end_2_end}.
For $\hat{a}\!<\!\hat{a}_c$, we obtain a single end-to-end distance in the two directions, and
at $\hat{a}\!=\!\hat{a}_c$, there is a discontinuous jump associated with elongation along the parallel direction
and retraction along the perpendicular direction.
With increasing field strength, 
the polymers become more elongated along the nematic director direction, and perpendicular fluctuations monotonically decrease.
The limit of high strength of alignment 
corresponds to a rigid rod-like state, where the end-to-end distance in the director direction approaches the contour length of the polymer, \emph{i.e.} $\langle R_{\parallel}^2\rangle \to L^2$. 

\begin{figure}[b]
    \centering
    \includegraphics[scale=0.4]{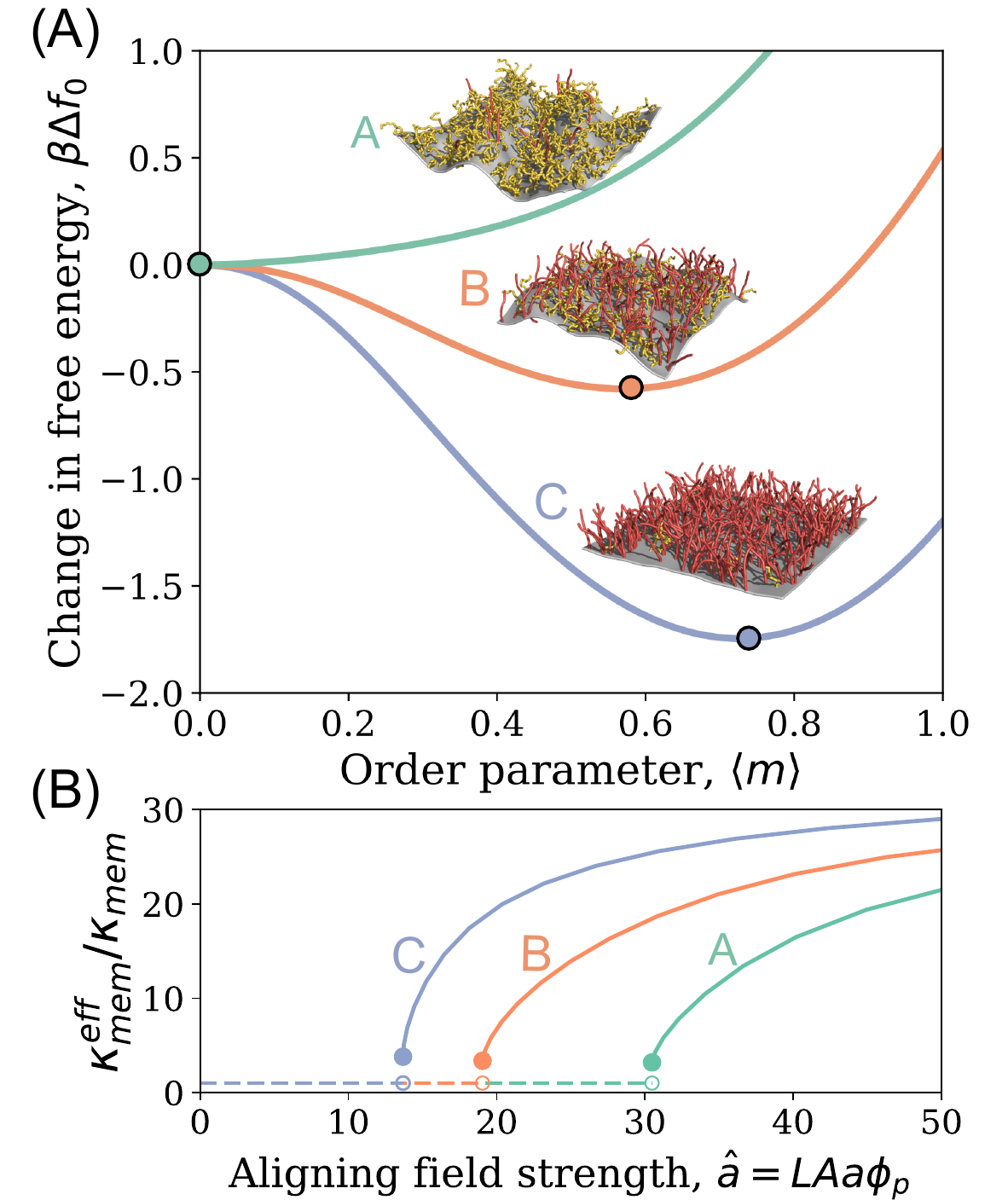}
    \caption{The top plot (A) shows the change in free energy with respect to the isotropic state is shown as a function of average order parameter variation for the three states of Fig.~\ref{fig:phase_diag} corresponding to rigid-polymer fractions of $0.04$, $0.50$ and $0.96$ ($A \to B \to C$) 
    for a fixed $\hat{a}=LAa\phi_p=27$. The bottom plot (B) shows the ratio of effective membrane rigidity as a function of aligning field strength for the three states.}
    \label{fig:free_energy}
\end{figure}
Within the mean-field approximation, we now consider free energies of the system as given by the purely nematic energy density in Eq.~\ref{Eq.free_energy}.
The order parameter of the entire system can be given as $\langle m\rangle$, and variational optimization of the free-energy density as a function of the order parameter gives the optimal value of $\langle m\rangle$ for a particular state.
We plot the excess nematic free energy density in Fig.~\ref{fig:free_energy}A for the three defined states for a fixed $\hat{a}=27$.
As we note from Fig.~\ref{fig:phase_diag}, state $A$ has $\hat{a}_c>27$, 
meaning the system would be in isotropic state and variational minimization of free energy predicts 
the system resides at $\langle m\rangle_0\!=\!0$, as noted by the circle.
For both state B and C, Fig.~\ref{fig:phase_diag} states $\hat{a}$ should produce a nematic state ($\langle m\rangle_0\!>\!0$), which is shown in Fig.~\ref{fig:free_energy}.
We note that as the fraction of rigid polymers increase from $A\to B\to C$, $\langle m\rangle_0$ monotonically grows from the isotropic to the nematic state towards $\langle m\rangle_0\!\to \!1$ indicating perfectly aligned polymer configurations.
As the fraction of rigid polymers increase, the nematic free energy of the system decreases indicating stabilizing forces in the system.
With the same surface coverage of EEA1 brush-like polymers on a membrane patch, we anticipate 
a decreasing free energy as the brushes exist in the rigid state.

We assign values of the EEA1 contour length $L = 222$~nm, cross-sectional area $A = 3.1$~nm$^2$~\cite{singh2023two,dumas2001multivalent}, and
EEA1 volume fraction $\phi_{p}=0.01$~\cite{murray2016endosomal,ohya2009reconstitution}.
The bare membrane bending elasticity is set to $\kappa_{\text{mem}} = 5k_{B}T$, which is within the measured physiological 
range~\cite{faucon1989bending}.
These parameters permit the evaluation of the effective membrane rigidity for the three rigid-polymer fractions shown in 
Fig.~\ref{fig:free_energy}A.
Figure~\ref{fig:free_energy}B gives the ratio of the effective membrane rigidity to the bare-membrane rigidity 
$\kappa_\text{mem}^{\text{eff}}/\kappa_\text{mem}$ as a function of aligning field strength.
For $\hat{a}\!<\!\hat{a}_c$ in each state, the effective membrane elasticity $\kappa_\text{mem}^{\text{eff}}\!=\!\kappa_{\text{mem}}$, since the 
splay modulus $K_{\text{splay}}$ of the filament brush is zero in the isotropic state.
At the nematic transition $\hat{a}\!=\!\hat{a}_c$, the effective membrane elasticity undergoes a 
first-order transition associated with the sudden increase in the nematic ordering.
Further increase in the aligning field strength leads to a greater than 20-fold increase 
in the effective elasticity of the membrane and the EEA1 brush,
leading to an overall picture where filament alignment dramatically modulates the membrane elasticity.

Notably, this dramatic change in the effective membrane elasticity is enabled by the collective nematic transition. 
The persistence lengths of the rigid and flexible states have a ratio of $l_p^{\text{EEA1}}/l_p^{\text{EEA1(GTP)}}\!\approx\!3.3$,
indicating a limiting ability for a single filament to physically manipulate exterior cargo.
However, the collective elastic behavior of the EEA1 brush would facilitate large-scale property changes that 
facilitate large deformations of the membrane necessary for membrane fusion.


In conclusion, we provide a theoretical model for collective
transitions in the EEA1 brush on an endosomal membrane, and
our analyses reveal how large-scale membrane fluctuations that are essential for membrane fusion are controlled by subtle physical and structural changes of the molecular constituents.
Our work suggests a possible mechanism of membrane fusion that relies on membrane fluidity (or reduced mechanical rigidity), which is modulated by the brush-layer protein filaments.
Extended EEA1 molecules that behave as effective rod-like polymers, 
increases the effective elastic rigidity of membrane through their collective alignment within 
the protein brush.
Conformational changes of EEA1 molecules (from a rigid to a flexible, floppy state) help physical proximity due to less steric hindrance and reduces the 
mechanical rigidity, allowing fusion to take place.

Our theoretical model and subsequent analyses addresses the role the 
coil-coiled dimeric structure of EEA1 protein in the mechanics of the
rigid-flexible transition, kinetics of Rab5-GTP binding, ATP$\to$ADP hydrolysis and Rab5-GDP unbinding, and the molecular mechanisms that
are responsible for rigid-to-flexible conformational change.
Our results have broad impact beyond the membrane-fusion problem. 
The model demonstrates how biophysical processes that happen on large, collective 
length and time scales can originate from orchestrated chemical and physical events at much finer scales,
which is a general principle that underlies a range of critical life processes.

\noindent
\emph{Acknowledgements.} A.G. acknowledges funding support from the Human Frontier Science Program, Grant No. HFSP/REF RGP0019/20.
A.J.S. was supported by the NSF program Condensed Matter and Materials Theory, Grant No. 1855334.

\bibliography{apssamp}

\clearpage
\newpage
\newpage
\setcounter{equation}{0}
\setcounter{figure}{0}
\setcounter{table}{0}
\setcounter{page}{1}
\makeatletter
\renewcommand{\theequation}{S\arabic{equation}}
\renewcommand{\thefigure}{S\arabic{figure}}
\renewcommand{\thepage}{S\arabic{page}}

\newpage
\onecolumngrid
\begin{center}
    \Large{\textbf{Supplemental Information}\\
    to\\
    \textbf{Local Changes in Protein Filament Properties
Drive Large-Scale Membrane Transformations Involved in
Endosome Tethering and Fusion}}
\end{center}

\tableofcontents
\section{Detailed Derivation of the Polymer Field Theory of a three component polymer nematic solution}
\noindent
\subsection{System Definition}
\noindent We consider an incompressible polymer solution of total volume $V$, with $n_{s}$ solvent molecule with volume $v_{s}$, $n_{p}^A$ polymer chains with contour length $L_A$ and cross-sectional area $A_A$ and $n_{p}^B$ polymer chains with contour length $L_B$ and cross-sectional area $A_B$.
The polymer chains are modeled using the wormlike chain model,
which describes the polymer chains as inextensible elastic threads that are subjected to thermal fluctuations.
The polymer configuration of the $i_\alpha$th polymer of type $\alpha\in (A,B)$ at arclength position $s$ (note, $s=0$ at one end and $s=L_{\alpha}$ at the opposite end) 
is defined by the space curve $\vec{r}_{i}^\alpha(s)$, and the local tangent vector $\vec{u}_{i}^\alpha(s) = \partial \vec{r}_{i}^\alpha(s)/\partial s$
gives the orientation of the monomer segment at $s$.
Inextensibility is strictly enforced by ensuring that $|\vec{u}_{i}^\alpha(s)| = 1$ for all configurations of the system.
The position of the $j$th solvent molecule is given by $\vec{r}_{j}$.
The bending rigidity of the semiflexible polymers are given by their respective persistence lengths $l_{p}^\alpha$,
and we also express the chain length in dimensionless units as
the number of Kuhn lengths $N_\alpha = L_\alpha/(2 l_{p}^\alpha)$,
where the Kuhn length $b_\alpha = 2 l_{p}^\alpha$ gives the statistical segment length of a polymer.

\par 

\noindent The system energy includes contributions for polymer deformation, solvent-polymer mixing, and 
nematic alignment of polymer chains and is given by the total energy,
\begin{eqnarray}
    \beta E = \sum_{\alpha\in \{A,B\}} \left\{ 
        \sum_{i=1}^{n_p^\alpha} \frac{l_p^\alpha}{2} \int_0^{L_{\alpha}} ds \left(\frac{\partial \vec{u}_i^{\alpha}(s)}{\partial s}\right)^2
        +\chi_\alpha \int d\Vec{r}\, \hat{\phi}_s(\vec{r})\hat{\phi}_p^\alpha(\vec{r})
        -\frac{a_\alpha}{2}\int d\Vec{r}\, \hat{\mathbf{S}}_\alpha(\vec{r}):\hat{\mathbf{S}}_\alpha(\vec{r}) \right\}\notag \\
        +\chi_{AB} \int d\Vec{r}\, \hat{\phi}_p^A(\vec{r})\hat{\phi}_p^B(\vec{r})
        -a_{AB} \int d\Vec{r}\, \hat{\mathbf{S}}_A(\vec{r}):\hat{\mathbf{S}}_B(\vec{r})
\end{eqnarray}
\noindent
where, $\hat{\phi}_{s}$ and $\hat{\phi}_{p}^\alpha$ are respectively the local dimensionless
density (volume fraction) of the solvent and polymer molecules of type $\alpha$, and $\hat{\mathbf{S}}_\alpha$
is the tensorial nematic order parameter density of state $\alpha$.
These local order parameters are given by,
\begin{eqnarray}
    \hat{\phi}_s(\vec{r}) =  v_s\sum_{j=1}^{n_s} \delta(\Vec{r}-\Vec{r}_j) \\
    \hat{\phi}_p^\alpha(\vec{r})  = A_\alpha \sum_{i=1}^{n_p^\alpha}  \int_0^{L_\alpha} ds \delta(\vec{r}-\vec{r}_i^{\alpha}(s))  \\
    \hat{S}_\alpha(\vec{r})  = A_\alpha \sum_{i=1}^{n_p^\alpha}  \int_0^{L_\alpha} ds \delta(\vec{r}-\vec{r}_i^{\alpha}(s))\left(\vec{u}_{i}^\alpha (s)\vec{u}_{i}^\alpha (s)-\frac{\mathbf{I}}{3}\right)
\end{eqnarray}
\noindent
For the subsequent analyses, we define a generalized volume fraction (tensor) that incorporates local orientational order, defined as~\cite{ghosh2022semiflexible},
\begin{equation}
    \hat{\phi}_{l_\alpha}^{m_\alpha} (\vec{r} \,) = A_\alpha
    \sqrt{\frac{4\pi}{2l_\alpha+1}}
    \sum_{i=1}^{n_{p}^\alpha}
    \int_{0}^{L_\alpha} \! \! \! d s
    \mathcal{Y}_{l_\alpha}^{m_\alpha}(\vec{u}_{i}^\alpha(s))
    \delta (\vec{r}-\vec{r}_{i}^\alpha(s)).
\end{equation}
$\mathcal{Y}_{l_\alpha}^{m_\alpha}$ is the real-valued
spherical harmonic (i.e. the tesseral spherical harmonic) for state $\alpha$, given by,
\begin{equation}
\mathcal{Y}_{l_\alpha}^{m_\alpha}=\begin{cases}
\sqrt{2} (-1)^{m_\alpha} \mathrm{Im}\left[ Y_{l_\alpha}^{|m_\alpha|} \right]
& \mathrm{for} \, \, \, m_\alpha<0\\
Y_{l_\alpha}^{0} & \mathrm{for} \, \, \,  m_\alpha=0\\
\sqrt{2} (-1)^{m_\alpha} \mathrm{Re}\left[ Y_{l_\alpha}^{m_\alpha} \right]
& \mathrm{for} \, \, \, m_\alpha>0
\end{cases}
\end{equation}
\noindent
where
$Y_{l_\alpha}^{m_\alpha}$ is the standard spherical harmonic (complex valued).
We note that the polymer volume fraction $\hat{\phi}_{p}^\alpha = \hat{\phi}_{0}^{0_\alpha}$, and
the nematic order parameter $\hat{\mathbf{S}}_\alpha$ can be written in terms of $\hat{\phi}_{2}^{m_\alpha}$.
With this definition, we rewrite the system energy as
\begin{eqnarray}
\beta E & = & 
\sum_{\alpha\in(A,B)}\left\{\sum_{i=1}^{n_{p}^\alpha} \frac{l_{p}^\alpha}{2} \int_{0}^{L_\alpha} 
\left( \frac{\partial \vec{u}_{i}^\alpha}{\partial s} \right)^{\! \! 2} d s +
\chi_\alpha \int d \vec{r} \, \hat{\phi}_{s}(\vec{r} \,) \hat{\phi}_{0}^{0_\alpha}(\vec{r} \,)
- \frac{a_\alpha}{3} \int d \vec{r} \, \sum_{m_\alpha = -2}^{2}  \left[ \hat{\phi}_{2}^{m_\alpha} (\vec{r} \,) \right]^{2}\right\} \nonumber\\
&  &   +\chi_{AB}\int d \vec{r} \, \hat{\phi}_{0}^{0_A}(\vec{r} \,) \hat{\phi}_{0}^{0_B}(\vec{r} \,) - \frac{2 a_{AB}}{3} \int d \vec{r} \, \sum_{m_A = -2}^{2}\sum_{m_B = -2}^{2}  \hat{\phi}_{2}^{m_A} (\vec{r} \,) \hat{\phi}_{2}^{m_B} (\vec{r} \,)\delta_{m_A m_B}
\label{eq:system_energy}
\end{eqnarray}
\noindent where the integrand in the last term in Eq.~\ref{eq:system_energy} signifies summation over terms such as, $\hat{\phi}_{2}^{-2_A} (\vec{r} \,) \hat{\phi}_{2}^{-2_B} (\vec{r} \,), \,  \hat{\phi}_{2}^{-1_A} (\vec{r} \,) \hat{\phi}_{2}^{-1_B} (\vec{r} \,), \dots$ and so on.

\noindent
The canonical partition function, $\mathcal{Z}$ is written as,
\begin{eqnarray}
    \mathcal{Z} = \frac{1}{n_s! n_p^A! n_p^B!}\frac{1}{v_s^{n_s}}\frac{1}{(L_A A_A)^{n_p^A} (L_B A_B)^{n_p^B} } \int \prod_{j=1}^{n_s}d\vec{r}_j \int \prod_{i_A=1}^{n_p^A} \mathcal{D}[\Vec{r}_{i_A}] \int \prod_{i_B=1}^{n_p^B} \mathcal{D}[\Vec{r}_{i_B}]\prod_{\Vec{r}} \delta(\hat{\phi}_s+\hat{\phi}_0^{0_A}+\hat{\phi}_0^{0_B}-1)\nonumber \\
    \times \prod_{i_A=1}^{n_p^A}\prod_{i_B=1}^{n_p^B}\prod_{s_A,s_B} \delta(|\partial_{s_A}\vec{r}_{i_A}|-1)\delta(|\partial_{s_B}\vec{r}_{i_B}|-1)\exp{(-\beta E)}
\end{eqnarray}
\noindent where $\displaystyle \prod_{\vec{r}} \delta(\hat{\phi}_s+\hat{\phi}_0^{0_A}+\hat{\phi}_0^{0_B}-1)$ accounts for the incompressibility constraint $\hat{\phi}_s+\hat{\phi}_0^{0_A}+\hat{\phi}_0^{0_B}=1$ at all locations within the system. 
The inextensibility of each polymer chain is enforced by the constraint.
The integration over $\mathcal{D}[\vec{r}_{i\alpha}]$ implies path integration over all conformations of the polymer $\alpha$.
We use the solvent volume $v_{s}$ and polymer volume $L_\alpha A_\alpha$  as volume scales instead of the  de-Broglie wavelengths cubed, which merely shifts the chemical potential by a constant quantity and does not affect the thermodynamic behavior of our system.

\subsection{Particle to Field Transformation}
\noindent
Next, we perform a series of field manipulations that enable systematic approximation of the thermodynamic behavior.
For each instantaneous volume fraction $\hat{\phi}$, we introduce a volume-fraction field variable by noting the 
property\cite{spakowitz2004exact,ghosh2022semiflexible}
\begin{eqnarray}
f[\hat{\phi}] & = & 
\int \mathcal{D} \phi \prod_{\vec{r}} \delta 
\!
\left[
\phi - \hat{\phi}
\right]f [\phi] \nonumber
\\
& = &
\int \mathcal{D} W \mathcal{D} \phi \exp \left\{ 
i \int d\vec{r} \, W(\vec{r} \,) \left[
\hat{\phi}(\vec{r} \,) - \phi(\vec{r} \,)
\right] 
\right\} f[\phi]
\end{eqnarray}
\noindent
where $W$ emerges from a Fourier representation of the spatial delta function.
This field manipulation is applied to each volume-fraction ($\hat{\phi}_{s}$, $\hat{\phi}_{0}^{0_\alpha}$, and $\hat{\phi}_{2}^{m_\alpha}$), resulting in the introduction of the conjugate fields $W_{s}$, $W_{0}^{0_\alpha}$, and $W_{2}^{m_\alpha}$.
After these field manipulations, we can write the canonical partition function as,
\begin{eqnarray}
\mathcal{Z} = \int 
\mathcal{D} W_{s}
\prod_{\alpha\in (A,B)}\left\{
\mathcal{D} \phi_{0}^{0_\alpha}
\mathcal{D} W_{0}^{0_\alpha}
\prod_{m_\alpha = -2}^{2}
\mathcal{D} W_{2}^{m_\alpha} 
\mathcal{D} \phi_{2}^{m_\alpha} \right\} 
\exp \Bigg\{ 
i \int d \vec{r} \, W_{s} (1-\phi_{0}^{0_A}-\phi_{0}^{0_B}) + 
i  \int d \vec{r}\sum_{\alpha\in(A,B)} W_{0}^{0_\alpha} \phi_{0}^{0_\alpha} 
\nonumber \\
+i \int d \vec{r}\sum_{\alpha\in(A,B)}
\sum_{m_\alpha=-2}^{2} 
W_{2}^{m_\alpha}
\phi_{2}^{m_\alpha} 
- \sum_{\alpha\in(A,B)}\chi_\alpha   \int d \vec{r} \,  \phi_{0}^{0_\alpha} (1-\phi_{0}^{0_A}-\phi_{0}^{0_B})
+\sum_{\alpha\in(A,B)} \frac{a_\alpha}{3} \int d \vec{r} \, \sum_{m_\alpha=-2}^{2} \left( \phi_{2}^{m_\alpha} \right)^{2}
\nonumber \\
- \chi_{AB}   \int d \vec{r} \,  \phi_{0}^{0_A}\phi_{0}^{0_B} +\frac{2 a_{AB}}{3} \int d \vec{r} \, \sum_{m=-2}^{2}\phi_{2}^{m_A}\phi_{2}^{m_B} +  
{n_{s}} \ln{\left(\frac{z_{s}[W_{s}]}{n_s v_s}\right)} + 
\sum_{\alpha\in(A,B)} n_p^\alpha \ln{\left(\frac{z_{p}^\alpha[W_{0}^\alpha,  W_2^{m_\alpha}]}{n_p^\alpha L_\alpha A_\alpha}\right)} \Bigg\}  \nonumber \\
 & & \hspace{-6.6 in}= \int 
\mathcal{D} W_{s}
\prod_{\alpha\in (A,B)}\left\{
\mathcal{D} \phi_{0}^{0_\alpha}
\mathcal{D} W_{0}^{0_\alpha}
\prod_{m_\alpha = -2}^{2}
\mathcal{D} W_{2}^{m_\alpha} 
\mathcal{D} \phi_{2}^{m_\alpha} \right\} \exp\left[-\beta \mathcal{F}\right] 
\label{eq:canon} 
\end{eqnarray}
\noindent
where, we define, the single solvent partition function as,
\begin{equation}
    z_{s}[W_{s}] = \int d \vec{r} \, \exp \left[
-i v_{s} W_{s}(\vec{r} \, )
\right]
\end{equation}
and the single polymer partition functions are written as,
\begin{eqnarray}
&   & z_{p}^\alpha[W_{0}^{0_\alpha}, W_{2}^{m_\alpha}] = \int \mathcal{D}[\vec{r} \,(s)] 
\exp \left\{
- \frac{l_{p}^\alpha}{2} \int_{0}^{L_\alpha} 
\left( \frac{\partial \vec{u}_\alpha}{\partial s} \right)^{\! \! 2} d s 
 - i A_\alpha \int_{0}^{L_\alpha} \! \! \! d s W_{0}^{0_\alpha}(\vec{r} \,(s)) 
 \right.
 \nonumber 
 \\
 &  &  \hspace{2 in}
 \left.
 - i A_\alpha \int_{0}^{L_\alpha} \! \! \!  d s \sum_{m_\alpha=-2}^{2} W_{2}^{m_\alpha}(\vec{r} \,(s)) \sqrt{\frac{4 \pi}{5}} \mathcal{Y}_{l_\alpha}^{m_\alpha}(\vec{u_\alpha}(s))
\right\}
\end{eqnarray}

\noindent The partition function $\mathcal{Z}$ has not been subjected to any approximations so far and is not exactly solvable as it is, although it is amenable to systematic approximations.
Below we formulate the general saddle-point approximation, which determines the thermodynamic state within a mean-field (generally inhomogeneous) and specialize to the homogeneous mean-field approximation.

\noindent
\section{Mean Field Approximation}
\subsection{Mean Field Equations}
\noindent
The lowest-order approximation to the canonical partition function is given by the saddle point, which approximates the functional integrals in Eq.~\ref{eq:canon} by the maximum term.
By setting the first variation of the argument of the exponential within Eq.~\ref{eq:canon} to zero (i.e. $\displaystyle \frac{\delta (\beta \mathcal{F})}{\delta f} = 0$ for any arbitrary field $f$), we arrive at the 
saddle-point equations.
Saddle point equations are given as (we use $\bar{\phi}_0^{0_\alpha}=\phi_p^\alpha$, and $iW=w$),
\begin{equation}
    (1-\phi_p^A-\phi_p^B) = \frac{v_s n_s}{\bar{z}_s}\exp{(-v_s\bar{w}_s)},
\end{equation}
\begin{equation}
    -\chi_A(1-2\phi_p^A-\phi_p^B)+\chi_B\phi_p^B+\bar{w}_0^{0_A}-\bar{w}_s-\chi_{AB}\phi_p^B = 0,
\end{equation}
\begin{equation}
    -\chi_B(1-\phi_p^A-2\phi_p^B)+\chi_A\phi_p^A+\bar{w}_0^{0_B}-\bar{w}_s-\chi_{AB}\phi_p^A = 0,
\end{equation}
\begin{equation}
    \frac{2a_A}{3}\Bar{\phi}_2^{m_A}+ \frac{2a_{AB}}{3}\Bar{\phi}_2^{0_B}+\Bar{w}_2^{m_A}=0
\end{equation}
\begin{equation}
    \frac{2a_B}{3}\Bar{\phi}_2^{m_B}+ \frac{2a_{AB}}{3}\Bar{\phi}_2^{m_A} + \Bar{w}_2^{m_B}=0
\end{equation}
\begin{equation}
    \bar{\phi}_p^A+ \frac{n_p^A}{\Bar{z}_p^A}\frac{\delta \bar{z}_p^A}{\delta \Bar{w}_0^{0_A}} = 0
\end{equation}
\begin{equation}
    \bar{\phi}_p^B+ \frac{n_p^B}{\Bar{z}_p^B}\frac{\delta \bar{z}_p^B}{\delta \Bar{w}_0^{0_B}} = 0
\end{equation}
\begin{equation}
    \bar{\phi}_2^{m_A}+ \frac{n_p^A}{\Bar{z}_p^A}\frac{\delta \bar{z}_p^A}{\delta \Bar{w}_2^{m_A}} = 0
\end{equation}
\begin{equation}
    \bar{\phi}_2^{m_B}+ \frac{n_p^B}{\Bar{z}_p^B}\frac{\delta \bar{z}_p^B}{\delta \Bar{w}_2^{m_B}} = 0
\end{equation}
Next, we specialize our treatment of the saddle point to the homogeneous uniaxial nematic state with nematic alignment along $\hat{\delta}_z$.
This gives $\Bar{\phi}_2^{m_\alpha}=0$ for $m_\alpha\ne 0$ and $\alpha\in (A,B)$.

\noindent
This leads to the self-consistent mean-field equation, (for $\alpha\in \{A,B\}$)
\begin{equation}
    \Bar{\phi}_2^{0_\alpha}=\frac{\phi_p^\alpha}{L_\alpha}\int_0^{L_\alpha} ds \left\langle \left(\frac{3}{2}[u_z^\alpha(s)]^2-\frac{1}{2}\right) \right\rangle_{0_\alpha} = \phi_p^\alpha m_N^\alpha 
\end{equation}
where the average $\langle \dots \rangle_{0_\alpha}$ is taken with respect to the single chain mean-field energy,
\begin{equation}
    \beta \mathcal{E}_0^\alpha =\frac{l_p^\alpha}{2}\int_0^{L_\alpha} ds \left(\frac{\partial \vec{u}^\alpha(s) }{\partial s}\right)^2 - (a_\alpha \phi_p^\alpha m_N^\alpha A_{\alpha}+a_{AB} \phi_p^\beta m_N^\beta A_\beta) \int_0^{L_\alpha} ds \left\{[u_z^\alpha(s)]^2-\frac{1}{3}\right\}
\end{equation}

\noindent
It is understood that $0\le m_N^\alpha \le 1, \forall \alpha\in\{A,B\}$.
The overall nematic order of the system can be thought of as an weighted average as, $\displaystyle m_{\text{total}}=\frac{1}{\phi_p}\sum_\alpha \phi_p^\alpha m_N^\alpha$.
For a perfectly aligned state ($m_N^\alpha=1, \forall \alpha$) we get, $\displaystyle \bar{\phi}_2^0 = \sum_\alpha \bar{\phi}_2^{0_\alpha}=\sum_{\alpha}\phi_p^\alpha=\phi_{p}$.
\subsection{Homogeneous Mean-field Approximation and System Energy}
\noindent
Under homogeneous mean-field approximation the zeroth order energy becomes (per unit volume $V$, so this is technically an energy density),
\begin{eqnarray}
    \beta f_0^{\text{iso.nem.}} = \frac{1-\phi_p}{v_s}\ln{(1-\phi_p)}+\sum_\alpha \frac{\phi_p^\alpha}{L_\alpha A_\alpha}\ln{(\phi_p^\alpha)}+\sum_\alpha \chi_\alpha \phi_p^\alpha(1-\phi_p)-\sum_\alpha \frac{a_\alpha}{3} (\bar{\phi}_2^{0_\alpha})^2 + \sum_{\alpha} \frac{\phi_p^\alpha}{L_\alpha A_\alpha}\ln{(q_\alpha)} \nonumber\\
     + \chi_{AB} \phi_p^A \phi_p^B - \frac{2a_{AB}}{3}\Bar{\phi}_2^{0_A}\Bar{\phi}_2^{0_B}
\end{eqnarray}
\noindent where $f_0 = \mathcal{F}_0/V$ and an intensive property of the system.
Here $q_\alpha$ is the partition function given as,
\begin{equation}
    \begin{split}
        q_\alpha = \int d\vec{u} \int d\vec{u}_0 \int_{\Vec{u}_0}^{\Vec{u}}\mathcal{D}[\vec{u}(s)] \exp{\left( -\beta \mathcal{E}_0^{\alpha}[\vec{u}(s)]\right)}
    \end{split}
\end{equation}
To note,
\begin{eqnarray}
    \beta f_0^{\text{iso.}} = \frac{1-\phi_p}{v_s}\ln{(1-\phi_p)}+\sum_\alpha \frac{\phi_p^\alpha}{L_\alpha A_\alpha}\ln{(\phi_p^\alpha)}+\sum_\alpha \chi_\alpha \phi_p^\alpha(1-\phi_p)
     + \chi_{AB} \phi_p^A \phi_p^B 
\end{eqnarray}
\noindent Here, it should be noted that in absence of any aligning field (\emph{i.e.} $\gamma_\alpha = (A_\alpha a_\alpha m_N^\alpha\phi_p^\alpha+A_\beta a_{AB} m_N^\beta\phi_p^\beta)$), the orientational partition is unity. Below we simplify to the notation $m_N^\alpha\equiv m_\alpha$.

\noindent The residual Helmholtz free energy (density) of the nematic phase relative to the isotropic phase is given by,
\begin{equation}
    \beta f_{\text{residual}} =\beta f_0^{\text{iso.}} -\beta f_0^{\text{iso.nem.}} = \sum_{\alpha} \left( \frac{a_\alpha}{3}(\phi_p^\alpha m_\alpha)^2 - \frac{\phi_p^\alpha}{L_\alpha A_\alpha}\ln{\left\{q_\alpha (m_\alpha)\right\}}\right) + \frac{2a_{\alpha\beta}}{3}\phi_p^\alpha m_\alpha \phi_p^\beta m_\beta
\end{equation}
For a constant $a_\alpha$, constant $L_\alpha=L$ and $A_\alpha=A$, we get the free energy (by multiplying with the relevant volume $LA/\phi_p$) as written in the main manuscript,
\begin{equation}
    \beta \Delta f_0 = \frac{\hat{a}}{3}\left(f_Rm_R+f_Fm_F\right)^2 - f_R\ln{q_R} - f_F\ln{q_F}
\end{equation}
where, $\hat{a}=a\phi_p L A$ and $f_\alpha = \phi_p^\alpha/\phi_p$. 
In the limit of $f_R\to 1$ (hence, $f_F\to 0$) the above equation reduces to (superscript `1' represents single type of polymer component),
\begin{equation}
    \beta \Delta f_0 ^{(1)} = \frac{\hat{a}}{3} m^2 - \ln{q}
\end{equation}
that matches with ref.~\cite{spakowitz2004exact}.

\noindent

\section{Fluctuations around mean field solution}

\noindent
Now we specialize to fluctuation effects upto quartic order with respect to the homogeneous mean field basis.
The vector of auxiliary variables are defined as, $\mathcal{W} \equiv [ W_s, \{ \phi_0^{0_\alpha}\}, \{ W_0^{0_\alpha}\}, \{ \phi_2^{m_\alpha}\}, \{ W_2^{m_\alpha}\} ]$ (total of $25$ terms) and the vector containing all the fluctuations are defined using, $\Delta = [\{ \delta\phi_0^{0_\alpha}\}, \{ \delta W_0^{0_\alpha}\}, \{ \delta \phi_2^{m_\alpha}\}, \{\delta W_2^{m_\alpha}\}]$ (total of $24$ terms due to the incompressibility constraint).
Performing Gaussian integral over the fluctuating conjugate fields (not volume fraction fluctuation fields), we arrive at,
\begin{equation}
    -\beta\mathcal{F} =  -\beta \mathcal{F}_0 - \frac{1}{2}\Tilde{\Gamma}_{\Tilde{1}\Tilde{2}}^{(2,\phi)}\Delta_{\Tilde{1}}\Delta_{\Tilde{2}} + \sum_{n=3}^{\infty} \frac{1}{n!}\Tilde{\Gamma}_{\Tilde{1}\dots\Tilde{n}}^{(n,\phi)}\Delta_{\Tilde{1}}\dots\Delta_{\Tilde{n}}
\end{equation}
\noindent where the above expansion is formally exact and written using modified Einstein notation to sum over fluctuating volume fraction fields of all constituents and integration over the respective Fourier variables.
The quadratic order term can be explicitly written as,
\begin{equation}
    \Tilde{\Gamma}_{\Tilde{1}\Tilde{2}}^{(2,\phi)}\Delta_{\Tilde{1}}\Delta_{\Tilde{2}} = \frac{1}{(2\pi)^6}\sum_{\gamma_1,\gamma_2}\int d\Vec{k}_1 d\Vec{k}_2 \Tilde{\Gamma}_{\gamma_1\gamma_2}^{(2,\phi)}(\Vec{k}_1,\Vec{k}_2)\Delta_{\gamma_1}(\Vec{k}_1)\Delta_{\gamma_2}(\Vec{k}_2)
\end{equation}
\noindent where, $\gamma$ runs over the three species present in the system, namely solvent, polymer $A$ and polymer $B$. In order to write the full matrix $\mathbf{\Gamma}^{(2,\phi)}$ in terms of its elements, we write the structure factors for polymer species $\alpha$ as,
\begin{eqnarray}
    S_{(l_1, m_1),(l_2,m_2)}^{(\alpha)}(\Vec{k}) & = & \frac{1}{L_\alpha^2}\left(\sqrt{\frac{4\pi}{(2l_1+1)(2l_2+1)}}\right)^{m_1+m_2} \nonumber\\
    & & \times \int_0^{L_\alpha}ds_1 \int_0^{L_\alpha} ds_2 \left\langle \mathcal{Y}_{l_1}^{m_1}(\Vec{u}(s_1)) \exp{\left[i\Vec{k}\cdot (\Vec{r}(s_1)-\Vec{r}(s_2)\right]}\mathcal{Y}_{l_2}^{m_2}(\Vec{u}(s_2))\right\rangle
\end{eqnarray}
\noindent The entire $\mathbf{\Gamma}^{(2,\phi)}$ matrix is a $12\times 12$ matrix containing all the volume fraction fluctuations. 
The matrix is sparse containing only certain non-zero elements ($\beta\ne \alpha$ in equations below and both indicates $A$ and $B$):
\begin{equation}
    \Gamma_{(0,0),(0,0)}^{(\alpha)} (\Vec{k}) = \frac{1}{v_s(1-\phi_p)} -2\chi_{\alpha} + \frac{1}{A_\alpha L_\alpha \phi_p^\alpha} [S_{(0,0),(0,0)}^{(\alpha)}]^{-1}(\vec{k})
\end{equation}
\begin{equation}
    \Gamma_{(0,0),(2,m)}^{(\alpha)} (\Vec{k}) = \frac{1}{A_\alpha L_\alpha \phi_p^\alpha} [S_{(0,0),(2,m)}^{(\alpha)}]^{-1}(\vec{k})
\end{equation}
\begin{equation}
    \Gamma_{(2,m_1),(2,m_2)}^{(\alpha)} (\Vec{k}) = -\frac{2a_\alpha}{3}\delta_{m_1m_2}+\frac{1}{A_\alpha L_\alpha \phi_p^\alpha} [S_{(2,m_1),(2,m_2)}^{(\alpha)}]^{-1}(\vec{k})
\end{equation}
\begin{equation}
    \Gamma_{(0,0),(0,0)}^{(\alpha\beta)} (\Vec{k}) = -2\chi_{\alpha\beta} 
\end{equation}
\begin{equation}
    \Gamma_{(2,m_1),(2,m_2)}^{(\alpha\beta)} (\Vec{k}) = -\frac{2a_{\alpha\beta}}{3}\delta_{m_1m_2} 
\end{equation}
\noindent To note, there is cross correlation of structure since $A$ and $B$ polymers are separate from one another (\emph{i.e} not a copolymer like structure).
The formulation above gives the exact free energy up-to quadratic order.

\noindent Next steps involve, (1) writing the quadratic order free energy within the RPA approximation in terms of the eigenvalues of $\mathbf{\Gamma}^{(2,\phi)}$ matrix, (2) defining a non-interacting reference free energy, (3) subtract this reference free energy and (4) using the property, $\displaystyle \prod_{\eta\in\{\text{fields}\}}\gamma_\eta=\text{det}[\mathbf{\Gamma}]$ (where, $\gamma$ are the eigenvalues of the determinant and the number of eigenvalues are same as the number of fluctuating volume fraction fields) to arrive at,
\begin{equation}
    \beta \mathcal{F} = \beta\mathcal{F}_0+\frac{V}{4\pi^2}\int_0^\Lambda dk \ k^2\ln{\left(\frac{\text{det}[\mathbf{\Gamma}^{(2,\phi)}]}{\text{det}[\mathbf{\Gamma}_0^{(2,\phi)}]}\right)} 
\end{equation}
\noindent In the above equation, we have introduced a $\text{high-}k$ cutoff, $\displaystyle \Lambda=\min_{\{l_p^\alpha\}} \frac{2\pi}{l_p^{\alpha}}$ to account for the ultraviolet divergence in field theory.

\section{Chemical potential and system Phase Behavior}
\noindent We only care about the phase behavior at the mean-field level.
Due to enforced incompressibility of the system the Gibbs and Helmholtz free energy of the system are equivalent to one other. 
Here, we define the chemical potential of $\eta^{\text{th}}$ species as,
\begin{equation}
    \beta\mu_\eta=\frac{\partial (\beta\mathcal{F})}{\partial n_{\eta}} = \beta v_\eta f+V\frac{\partial (\beta f)}{\partial n_\eta}
\end{equation}
where the second equality follows from defining the free energy density, as $f=\mathcal{F}/V$.
The expression for the zeroth order chemical potential of species $\theta$ follows from the expression of mean field free energy of the system.
We write the mean field energy in a consolidated notation as,
\begin{eqnarray}
    \beta f_0 = \sum_{k} \frac{\phi_k}{v_k}\ln{\left(\frac{\phi_k}{q_k}\right)} + \frac{1}{2} \sum_{j,k} \chi_{jk}\phi_j\phi_k -\frac{1}{3} \sum_{j,k}a_{jk}m_jm_k \phi_j\phi_k \\
    = \sum_{k} \frac{\phi_k}{v_k}\ln{\left(\frac{\phi_k}{q_k}\right)} + \frac{1}{2} \sum_{j,k} \Xi_{jk}\phi_j\phi_k
\end{eqnarray}
\noindent where $\displaystyle \Xi_{jk} = \chi_{jk}-\frac{2}{3}a_{jk} m_j m_k$.
This leads to the expression for mean field chemical potential for species $i$ as, (where, $i\in \{\text{solvent}, \text{polymer }A, \text{polymer }B\}$)
\begin{equation}
    \beta\mu_0^{(i)} = 1 + \ln{\left(\frac{\phi_i}{q_i}\right)} - v_i\sum_{j} \frac{1}{v_j}\phi_j + v_i\sum_{j} \Xi_{ij} \phi_j - \frac{v_i}{2}\sum_{jk} \Xi_{jk} \phi_j \phi_k
\end{equation}
\noindent where all the $\phi_i$ denote the mean-field volume fraction of species $i$.

\section{Effective Membrane Rigidity}
\subsection{Frank Elastic Energy of Polymer Brush Layer}

\begin{figure}[ht!]
    \centering
    \includegraphics[scale=0.40]{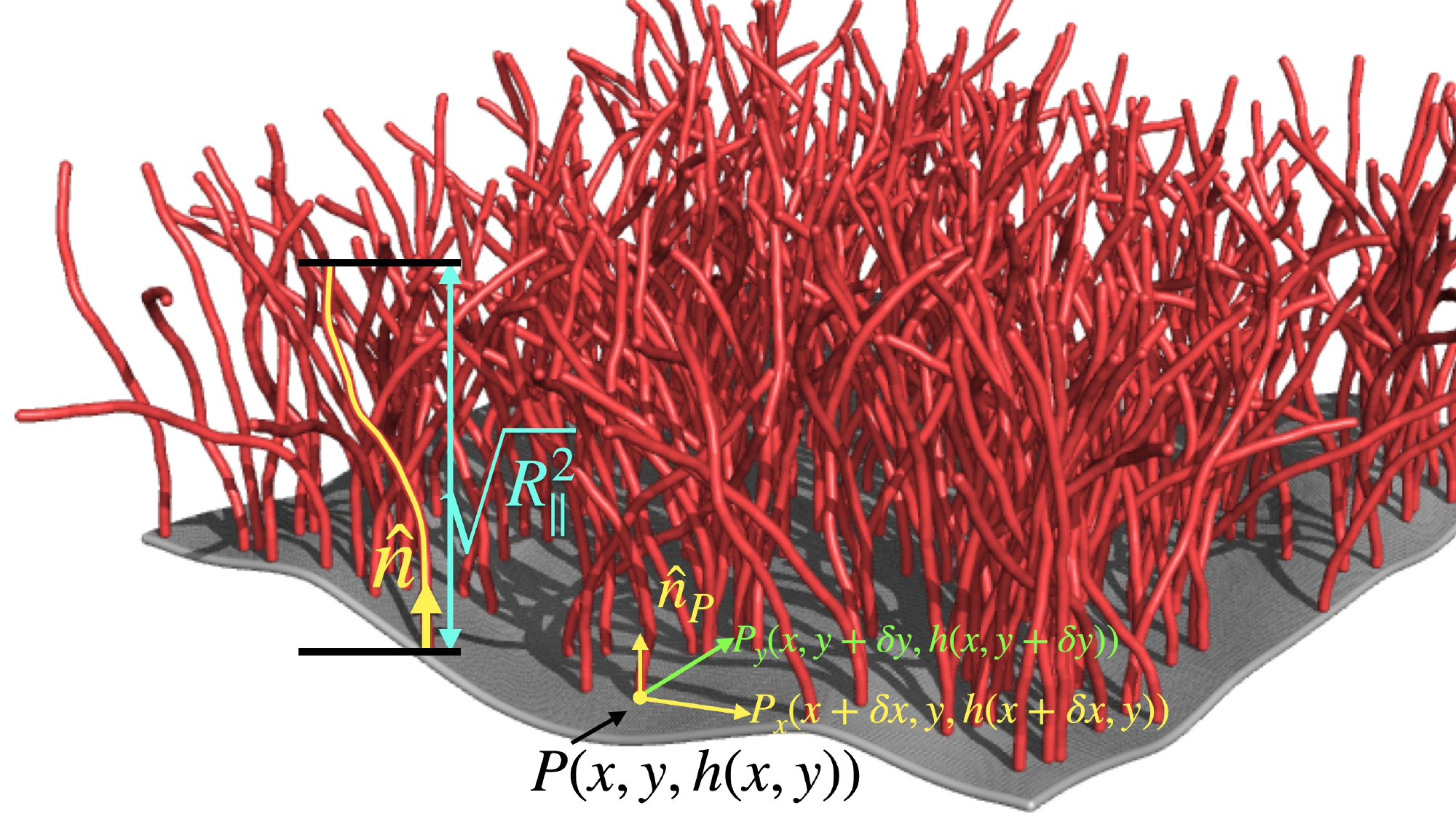}
    \caption{A membrane patch is shown to schematically show normal at any point on the membrane that holds a polymer. The polymer structural axis is shown in yellow solid line and the polymer end-to-end distance along the preferred direction of alignment in its extended state is shown in cyan double-headed arrow. Construction of unit normal $\hat{n}_P$ at point $P\equiv(x,y,h(x,y))$ is demonstrated as the normalized cross product of $\overrightarrow{PP}_x$ and $\overrightarrow{PP}_y$}
    \label{fig:schematic2}
\end{figure}

\noindent Here, we briefly describe how existence of polymer brush layer in its extended state is responsible for renormalizing the membrane bending rigidity.
Noting that one end of the polymer is tethered to the membrane, we approximate the Frank elastic energy of the polymer brush layer as follows~\cite{ghosh2022semiflexible},
\begin{eqnarray}
\beta F_{\mathrm{elas}} & = &  \frac{1}{2} \int d \vec{r} \,
\left\{
K_{\mathrm{bend}} \left( \vec{n} \times \vec{\nabla} \times \vec{n} \right)^{2}
+K_{\mathrm{twist}} \left( \vec{n} \cdot \vec{\nabla} \times \vec{n} \right)^{2}
+ K_{\mathrm{splay}} \left( \vec{\nabla} \cdot \vec{n} \right)^{2}
\right\} \\
& & \hspace{-0.1 in} \approx \frac{1}{2} 
\left\{
K_{\mathrm{bend}} \left( \vec{n}_P \times \vec{\nabla} \times \vec{n}_P \right)^{2}
+K_{\mathrm{twist}} \left( \vec{n}_P \cdot \vec{\nabla} \times \vec{n}_P \right)^{2}
+ K_{\mathrm{splay}} \left( \vec{\nabla} \cdot \vec{n}_P \right)^{2}
\right\}\sqrt{R_{\parallel}^2}
\nonumber
\end{eqnarray}
where $\vec{n}_P$ is the nematic director direction at polymer leg on the membrane (point $P$) and couples directly to membrane height fluctuation field.
Hence, $\vec{n}_P$ at point $P$ on the membrane can be identified as a normal to the membrane at $P$.
This can be calculated by parametrizing the membrane as $P\equiv (x,y,h(x,y))$, where $h(x,y)$ represents the height of the membrane at $2d$ grid point on the membrane $(x,y)$.
Hence, the normal at point $P$ is given as,
\begin{eqnarray}
    \vec{n}_P = \frac{\overrightarrow{PP}_x\times \overrightarrow{PP_y}}{||\overrightarrow{PP_x}\times \overrightarrow{PP_y}||} = \frac{-h_x\hat{\delta}_x-h_y\hat{\delta}_y+\hat{\delta}_z}{\sqrt{1+h_x^2+h_y^2}}\approx -h_x\hat{\delta}_x-h_y\hat{\delta}_y+\hat{\delta}_z
\end{eqnarray}
where, $\displaystyle h_x=\frac{\partial h(x,y)}{\partial x}$ and the approximate identity gives us variations with respect to minimal order changes in membrane height fluctuations.
Looking at the bend and twist terms of Frank elastic energy we note that, 
$\displaystyle (\vec{\nabla}\times \vec{n}_P) = h_{zy}\hat{\delta}_x-h_{zx}\hat{\delta}_y 
$, 
where $\displaystyle h_{zy}=\frac{\partial^2 h(x,y)}{\partial z \partial y}$ and thus are all higher order terms. 
This implies only splay rigidity of the polymers contribute to the Frank elastic energy to minimal order and is given as,
\begin{eqnarray}
    \beta F_{\text{elas}}  \approx \frac{1}{2}K_{\text{splay}} \sqrt{R_{\parallel}^2} \Big(h_{xx}+h_{yy}\Big)^2
\end{eqnarray}
\subsection{Helfrich Free Energy of Membrane and Renormalized Bending Rigidity}
For our case, we consider a membrane with the symmetric distribution of all membrane components (such as lipids and proteins) and the same environment on both sides of the membrane that may ensure symmetric membrane fluctuations and, thereby zero spontaneous curvature. 
The free energy for such a symmetric membrane is generally expressed as the Helfrich Hamiltonian~\cite{helfrich1973elastic}, given as,
\begin{flalign}
    \beta\mathcal{F}[h(\vec{x})] = \int d^2 {\vec{x}} \left\{ \frac{\nu_{\text{mem}}}{2}(\vec{\nabla} h(\vec{x}))^2 + \frac{\kappa_{\text{mem}}}{2}(\nabla^2 h(\vec{x}))^2  \right\} 
\end{flalign}
\noindent where, $h(\vec{x})$ describes the height of the membrane at location $\vec{x}\equiv (x,y)$ with respect to a reference state.
In the above equation $\nu_{\text{mem}}$ and $\kappa_{\text{mem}}$ represent the membrane tension and (bending) rigidity respectively.
The membrane bending rigidity at any point per above Helfrich Hamiltonian can be written as, $\displaystyle \frac{\kappa_{\text{mem}}}{2}(h_{xx}+h_{yy})^2$. 
Hence the membrane rigidity is renormalized due to Frank elastic energies of polymers and is given by,
\begin{eqnarray}
    \kappa_{\text{mem}}^{\text{eff}} = \kappa_{\text{mem}} + K_{\text{splay}}\sqrt{R_\parallel^2} \ge \kappa_{\text{mem}}
\end{eqnarray}
Qualitatively, we can infer in the isotropic state of polymer brush layer, the splay rigidity do not contribute to the overall membrane bending resistance.
However, in the nematic extended state of the polymer brush layer, we have finite $K_{\text{splay}}$ and effective membrane rigidity increases to larger stiffness prone to lesser fluctuations due to thermal and active forces in the medium.
In essence the ratio, $\Theta_{p/m}=K_{\text{splay}}\sqrt{R_\parallel^2}/\kappa_{\text{mem}}$ controls membrane fluctuations to zeroth order.

\section{Supplementary Figures}
\begin{figure}[h!]
    \centering
    \includegraphics[scale=0.60]{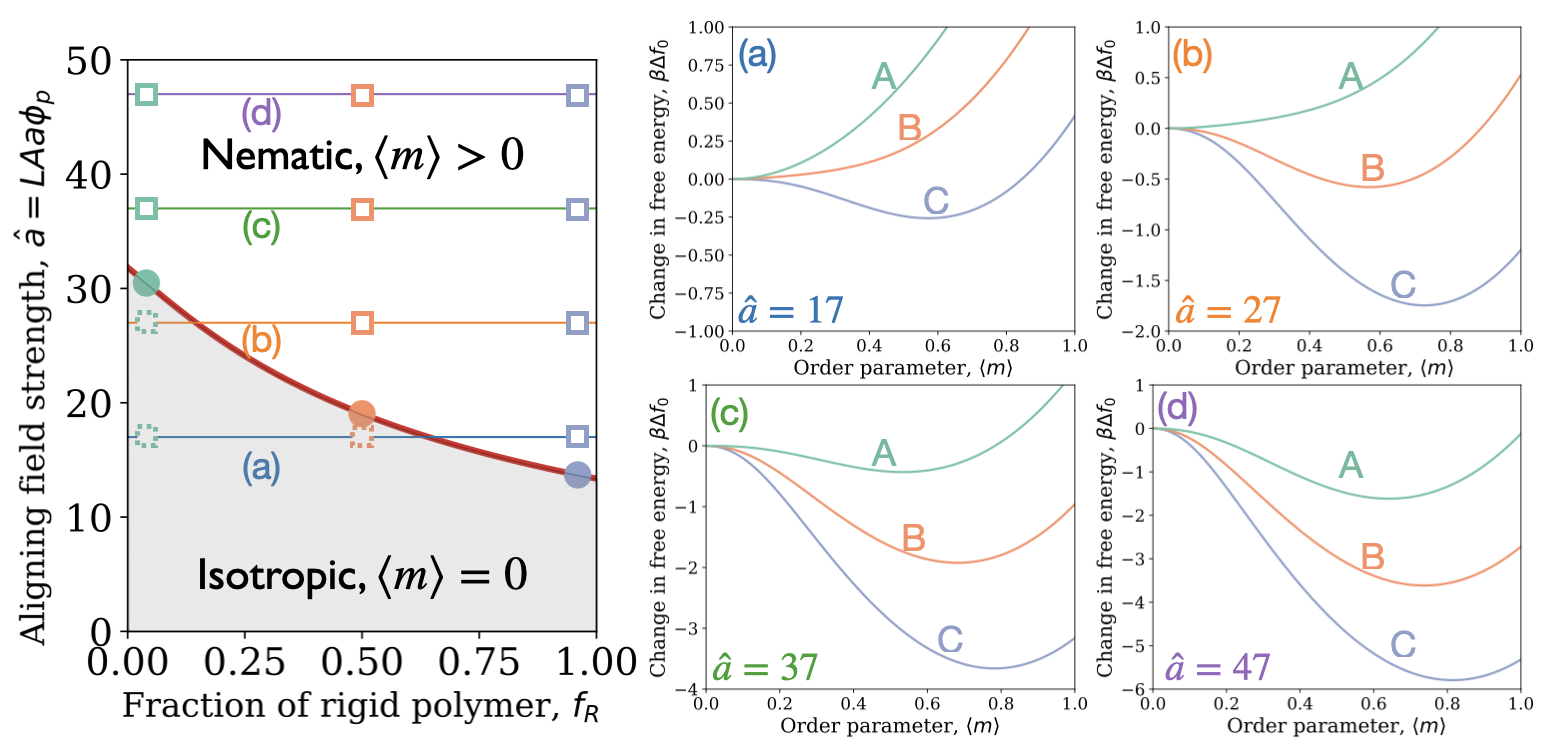}
    \caption{Phase diagram showing the boundary (thick solid line) of isotropic (gray) and nematic states as a function of rigid polymer fraction, $f_R$. Figures ((a)-(d)) show changes in free energy (cf. Fig.4 in main article) for state A ($f_R=0.04$), state B ($f_R=0.50$) and state C ($f_R=0.96$) for $\hat{a}=17, 27, 37, 47$ (panel (a) through (d) respectively). The states that are isotropic are designated as dashed squares in the left phase diagram while the states that shows nematic ordering are shown in solid squares in the phase diagram.}
    \label{fig:free_energy_supmat}
\end{figure}

\end{document}